\documentclass[aps,prd%,twocolumn
,preprint,tightenlines,superscriptaddress,nofootinbib,showpacs]{revtex4}
\usepackage{amssymb,latexsym}
\usepackage{amsmath,amsbsy,bbm}
\usepackage{epsfig,bm}
\usepackage{graphicx,comment}
\DeclareMathOperator{\tr}{tr}

\begin{document}
\def\a{{\alpha}}
\def\b{{\beta}}
\def\d{{\delta}}
\def\D{{\Delta}}
\def\e{{\varepsilon}}
\def\g{{\gamma}}
\def\G{{\Gamma}}
\def\k{{\kappa}}
\def\l{{\lambda}}
\def\L{{\Lambda}}
\def\m{{\mu}}
\def\n{{\nu}}
\def\o{{\omega}}
\def\O{{\Omega}}
\def\S{{\Sigma}}
\def\s{{\sigma}}
\def\th{{\theta}}

\def\ol#1{{\overline{#1}}}

\def\Dslash{D\hskip-0.65em /}
\def\Dtslash{\tilde{D} \hskip-0.65em /}

\def\CPT{{$\chi$PT}}
\def\QCPT{{Q$\chi$PT}}
\def\PQCPT{{PQ$\chi$PT}}
\def\tr{\text{tr}}
\def\str{\text{str}}
\def\diag{\text{diag}}
\def\order{{\mathcal O}}

\def\cC{{\mathcal C}}
\def\cB{{\mathcal B}}
\def\cT{{\mathcal T}}
\def\cQ{{\mathcal Q}}
\def\cL{{\mathcal L}}
\def\cO{{\mathcal O}}
\def\cA{{\mathcal A}}
\def\cQ{{\mathcal Q}}
\def\cR{{\mathcal R}}
\def\cH{{\mathcal H}}
\def\cW{{\mathcal W}}
\def\cM{{\mathcal M}}
\def\cD{{\mathcal D}}
\def\cN{{\mathcal N}}
\def\cP{{\mathcal P}}
\def\cK{{\mathcal K}}
\def\Qt{{\tilde{Q}}}
\def\Dt{{\tilde{D}}}
\def\St{{\tilde{\Sigma}}}
\def\cBt{{\tilde{\mathcal{B}}}}
\def\cDt{{\tilde{\mathcal{D}}}}
\def\cTt{{\tilde{\mathcal{T}}}}
\def\cMt{{\tilde{\mathcal{M}}}}
\def\At{{\tilde{A}}}
\def\cNt{{\tilde{\mathcal{N}}}}
\def\cOt{{\tilde{\mathcal{O}}}}
\def\cPt{{\tilde{\mathcal{P}}}}
\def\cI{{\mathcal{I}}}
\def\cJ{{\mathcal{J}}}

\def\eqref#1{{(\ref{#1})}}

\preprint{UMD-40762-400}
 
\title{Pion Polarizabilities and Volume Effects in Lattice QCD}

\author{Jie Hu}
\email[]{hujie@phy.duke.edu}
\affiliation{Department of Physics, Duke University, Box 90305, 
Durham, NC 27708-0305, USA}
\author{Fu-Jiun Jiang}
\email[]{fjjiang@itp.unibe.ch}
\affiliation{Institute for Theoretical Physics, Bern University, Sidlerstrasse 5, CH-3012 Bern, Switzerland}
\author{Brian C.~Tiburzi}
\email[]{bctiburz@umd.edu}
\affiliation{Department of Physics, Duke University, Box 90305, 
Durham, NC 27708-0305, USA}
\affiliation{Maryland Center for Fundamental Physics, Department of Physics, 
University of Maryland, College Park, MD 20742-4111, USA}

\date{\today}


\begin{abstract}
We use chiral perturbation theory to study the extraction of pion 
electromagnetic polarizabilities from lattice QCD. 
Chiral extrapolation formulae are derived for
partially quenched QCD, and quenched QCD simulations. 
On a torus, volume dependence of electromagnetic observables is complicated by 
$SO(4)$ 
breaking, as well as photon zero-mode interactions.
We determine finite volume corrections to the
Compton scattering tensor of pions. 
We argue, however, that such results cannot be used to
ascertain volume corrections to polarizabilities 
determined in lattice QCD with background field methods.
Connection is lacking because 
momentum expansions are not
permitted in finite volume.
Our argument also applies to form factors.
Volume effects for electromagnetic moments
cannot be deduced from finite volume form factors. 
\end{abstract}
\maketitle

\section{Introduction}

Electromagnetic polarizabilities
encode fundamental properties of bound states.
The electric polarizability of the ground state hydrogen
atom, for example,
$\alpha^{H}_{E}= \alpha_{fs} N / m_e E_0^2$, 
represents the ease at which the atomic electron cloud deforms in an 
applied electric field. 
Here 
$\alpha_{fs} = e^2 / 4 \pi$ 
is the fine structure constant, 
$m_e$ 
is the electron mass, 
$E_0$ 
is the ground state energy, and 
$N$ is a pure number, 
which turns out to be 
$9/8$.  Atomic polarizability data 
are well described by theoretical calculations 
using atomic wave-functions of the weakly bound electrons. 
Hadronic polarizabilities, on the other hand, involve non-perturbative physics. 
The electrically charged quarks inside hadrons respond to applied electromagnetic fields
but against the strong chromo-electromagnetic forces that confine them 
into bound states. If the pion were a weakly bound
system of quarks with mass $m_q$, 
we might expect its electric polariazability to be of the form, 
$\alpha_E^\pi \sim \alpha_{fs} N  / m_q  m_\pi^2$.
The actual behavior is considerably different,
$\alpha^\pi_E  = \alpha_{fs} N / m_\pi \Lambda_\chi^2$,
where 
$\Lambda_\chi$ is the chiral symmetry breaking scale. 
It thus appears that the pion cloud of the pion is what deforms in the applied 
field, and that the relevant energy scale is $\Lambda_\chi$, which is an order 
of magnitude greater than the pion mass. Compared to the weakly 
bound scenario, the electric polarizability is a few orders of magnitude smaller,
which indicates stiffness of quarks inside hadrons.

Chiral peturbation theory (\CPT)~%
\cite{Gasser:1983yg}
provides a low-energy effective theory of 
QCD from which the pion polarizabilities can be calculated in terms of a 
few low-energy constants~%
\cite{Holstein:1990qy}. 
At leading order in the chiral power counting, 
calculated values for the pure number 
$N$ 
are 
$N^{\pi^0} = - 1/3$ 
for the neutral pion, and 
$N^{\pi^\pm} \approx 1/6$ 
for the charged pions. 
Comparing these polarizability predictions to experimental data 
is unlike the situation with atomic polarizabilities.
Without stable targets, experimental determination
is considerably challenging at best.
Pion polarizabilities, however, have been probed indirectly
in several experiments. 
Three reactions are used: 
radiative pion-nucleon scattering 
($\pi N \to \pi N \gamma$), 
pion photoproduction in photon nucleus scattering
($\gamma A \to \gamma A \pi$), 
and pion production seen in electron-positron
collisions
($\gamma^* \gamma \to \pi \pi)$. 
Neutral pion polarizabilities have been 
accessed only by the last reaction  
by the Crystal Ball Collaboration~%
\cite{Marsiske:1990hx}.
The most recent experimental effort has been by MAMI at Mainz~%
\cite{Ahrens:2004mg}
in measuring the difference of electric and magnetic
polarizabilities of the charged pion
through radiative pion-nucleon scattering, 
and by Compass at CERN~%
\cite{Colantoni:2005ku}
to measure charged pion polarizabilities
using photoproduction off lead. 
In the latter experiment, final data are being taken,
and soon will be analyzed. 
After the $12 \, \texttt{GeV}$ upgrade, Jefferson
Lab has plans to measure pion polarizabilities in the future.

Experiments to determine pion polarizabilities
have one feature in common: disagreement
with predictions from chiral perturbation theory. 
Considerable effort has been expended to determine
polarizabilities to two-loop order in \CPT ~%
\cite{Burgi:1996mm,Burgi:1996qi,Gasser:2005ud,Gasser:2006qa},
but discrepancy with experiment remains. 
Because these experiments are indirect, the
challenge is removing the hadronic
background to isolate the signal. This is a largely
model-dependent process with uncontrolled
systematic error.  
Recent dispersion relation calculations, however,
appear consistent with experimental 
values for the polarizabilities~%
\cite{Fil'kov:2005yf,Fil'kov:2005ss}.
Thus it remains unclear whether 
disagreement between theory and 
experiment has its roots in the 
experimental analysis, or in the 
behavior of the chiral expansion.

As a first principles
method, lattice QCD~%
\cite{DeGrand:2006aa} 
can be employed to 
determine pion polarizabilities. Currently 
and foreseeably this is itself a considerable 
challenge, but progress has been made
with background field methods~%
\cite{Fucito:1982ff,Martinelli:1982cb,Bernard:1982yu,Fiebig:1988en,Burkardt:1996vb,Christensen:2004ca,Detmold:2004kw,Lee:2005dq,Detmold:2006vu,Engelhardt:2007ub}.
Such calculations suffer a number of 
systematic errors, such as: 
quenching or partial quenching, 
quenching of sea quark charges, 
and volume effects.
While predictions of physical polarizabilities
are not currently possible from lattice QCD alone,  
forthcoming lattice QCD data on polarizabilities 
can be used as a diagnostic for \CPT. 
The predictions of \CPT\ can be tested against
lattice QCD data.
To this end, we perform a one-loop 
analysis of the quenching and partial quenching effects,
as well as the volume dependence of pion Compton scattering. 
As polarizabilities are the coefficients at second order in an expansion 
in photon momentum $\omega$, one would naively expect
that finite volume corrections to polarizabilities can be determined
from momentum expanding the finite volume Compton tensor. 
We find this is not the case. There are many terms in the finite volume
Compton tensor not anticipated by infinite volume gauge invariance. 
All terms, moreover, are form factors in $\omega L$, where $L$
is the spatial size of the lattice. Because of momentum quantization, 
these form factors cannot be expanded in $\omega L$. 
Thus the infinite volume connection between the frequency 
expansion and the polarizabilities is lost.
As polarizabilities are typically calculated in lattice QCD using background field 
methods, this means we cannot use the finite volume
Compton tensor to deduce finite volume corrections to polarizabilities
extracted from background field correlation functions. 
The same problem exists for electromagnetic moments.
Their volume effects cannot be deduced from 
series expanding finite volume electromagnetic form factors
about zero momentum transfer.

Our work has the following organization. First in 
Section~\ref{pcs}, 
we detail our conventions
for Compton scattering and the electromagnetic
polarizabilities of the pion. In 
Section~\ref{pqcpt}, 
we review the 
low-energy effective theories of QCD, and partially 
quenched QCD. Quenched QCD is discussed in 
Appendix~\ref{quenched}. 
These theories are then utilized in 
Section~\ref{ppiv} 
to compute the pion 
electromagnetic polarizabilities in infinite volume. 
Next in 
Section~\ref{ppfv}, 
we consider the modifications
to polarizabilities in finite volume. 
These modifications are complicated by both 
$SO(4)$ 
breaking and photon zero-mode interactions. 
We determine the finite volume modifications to  
pion Compton scattering. Here we argue, however, that these
modifications cannot be straightforwardly utilized to ascertain
finite volume effects for background field
calculations of polarizabilities in lattice QCD. 
Section~\ref{summy} 
summarizes our work, 
and Appendix~\ref{fvf} collects the finite 
volume functions employed in the main text.

\section{Pion Compton Scattering}

\subsection{Compton Scattering Amplitude}\label{pcs}
For Compton scattering in infinite volume, 
the amplitude for a real photon
to scatter off a pion can be parametrized as
\begin{eqnarray} 
T_{\gamma \pi}	
&=& 
2 m_\pi 
\left[
\left(
- \frac{e^2 Q_\pi^2}{m_\pi}
+ 
4 \pi
\,  
\alpha_E
\,
\omega^2
\right)
\bm{\varepsilon}'^*	
\cdot 
\bm{\varepsilon}
+
4 \pi 
\,
\beta_M
\, 
\omega^2
(\bm{\varepsilon}'^* \times \hat{\bm{k}}')
\cdot 
(\bm{\varepsilon} \times \hat{\bm{k}})
\right]
\notag \\
&& 
+ 
\frac{e^2 Q_\pi^2}{2 m_\pi^2} \omega^2  
(\bm{\varepsilon}'^* \cdot \hat{\bm{k}}) 
(\bm{\varepsilon} \cdot \hat{\bm{k}}')
( 1 - \cos \theta)
+ 
\ldots
\label{compamp}
,\end{eqnarray}
where in the center-of-momentum frame the photon momenta are
$k_\mu = (\omega, \omega \hat{\bm{k}})$ 
for the initial state, and
$k'_\mu = (\omega, \omega \hat{\bm{k}}')$
for the final state.  Terms denoted by 
$\ldots$
are higher order in the photon energy.
The frame-dependent scattering angle 
$\theta$ 
is given by 
$\cos \theta = \hat{\bm{k}}' \cdot \hat{\bm{k}}$.
In the above expression, 
$Q_\pi$ 
is the charge of the pion in units of 
$e>0$. 
In writing the physical amplitude, we have made use of 
Coulomb gauge in which the initial and final polarization vectors,
$\varepsilon_\mu$ and $\varepsilon'^*_\nu$, 
satisfy 
$\varepsilon_0 = \varepsilon'^*_0 = 0$.
The Compton amplitude appearing above, moreover, includes the one-particle 
reducible and irreducible pieces, as we have retained the Born terms.

The frequency independent term proportional to 
$Q_\pi^2$ 
reproduces the Thomson cross-section for low-energy 
scattering of charged particles when the amplitude squared 
is combined with appropriate phase-space factors. 
This term is exactly fixed by the total charge of the system in accordance with 
the Gell-Mann--Golberger--Low low energy theorems~\cite{Gell-Mann:1954kc,Low:1954kd}.
The induced E1-E1 interaction strength 
$\alpha_E$ 
is the electric polarizability, while the induced  
M1-M1 interaction strength 
$\beta_M$ 
is the magnetic
polarizability. In order to identify these as polarizabilities
one must pull out a factor of twice the target mass from the
Compton amplitude, as we have in Eq.~\eqref{compamp}. 
The electric and magnetic polarizabilities are the first 
structure dependent terms in the low-energy expansion of 
the Compton scattering amplitude. 
These polarizabilities can be determined from first
principles using lattice QCD techniques. In order
to make the connection between lattice data and 
real world QCD, extrapolations in quark mass and lattice
volume are required. To perform these requisite extrapolations, 
we turn to the low-energy effective theory of QCD, \CPT.

\subsection{\PQCPT\ for Pion Compton Scattering} \label{pqcpt}

In current lattice calculations, valence and sea quarks are often treated differently.
In the rather extreme approximation known as quenched QCD, the sea quarks are completely
absent. Less extreme is partially quenched QCD, where sea quarks are retained but 
have different masses than their valence counterparts. While both approximations are 
certainly contrary to nature, the latter retains QCD as a limit.  Observables
computed in partially quenched QCD can be connected to their real world values
by utilizing partially quenched \CPT\ (\PQCPT) to derive formulae 
for the required extrapolation in sea quark mass.  
Because \CPT\ is contained as a limiting case of \PQCPT, we focus our discussion on \PQCPT. 
Peculiarities of quenched \CPT\  (\QCPT) will be noted where relevant
and the general conventions appear in Appendix~\ref{quenched}.
For further details on \QCPT\ and \PQCPT, 
see~\cite{Morel:1987xk,Sharpe:1992ft,Bernard:1992mk,Sharpe:1997by,Golterman:1998st,Sharpe:2000bc,Sharpe:2001fh}.

To determine pion observables, we imagine that the strange quark mass 
is fixed at the physical value so that no extrapolations are needed in 
the valence strange or sea strange quark masses. To this end, we consider
a partially quenched theory of valence $u$ and $d$ quarks, paired with degenerate
ghost quarks $\tilde{u}$ and $\tilde{d}$, and two additional sea quarks $j$ and $l$.
The quark masses are given in a matrix
\begin{equation}
m_Q = \diag \left( m_u, m_d, m_j, m_l, m_u, m_d \right)
,\end{equation} 
where the last two entries are ghost quark masses that are degenerate
with their valence counterparts.
For simplicity below, we work in the isospin limit of the valence and sea sectors, 
so that $m_u = m_d$ and $m_j = m_l$. 
\PQCPT\ describes the low-energy dynamics of partially quenched QCD
and is written in terms of the mesons $\Phi$
that appear in the coset field $\Sigma$ as\footnote{%
In our conventions, the pion decay constant $f \approx 132 \, \texttt{MeV}$.
}
\begin{equation}
\Sigma = \exp \left( \frac{2 i \Phi}{f} \right)
.\end{equation}
These mesons are the pseudo-Goldstone modes appearing from spontaneous
chiral symmetry breaking: $SU(4|2)_L \otimes SU(4|2)_R \to SU(4|2)_V$.
The dynamics of these modes is described at leading-order by 
the \PQCPT\ Lagrangian
\begin{equation} \label{eq:L}
\cL 
= 
\frac{f^2}{8} 
\str \left( D_\mu \Sigma^\dagger D^\mu \Sigma \right)
+
\lambda \frac{f^2}{4} 
\str \left( m_Q^\dagger \Sigma + \Sigma^\dagger m_Q \right)
- m_0^2 \Phi_0^2
.\end{equation}
Here $\Phi_0 = \str (\Phi) / \sqrt{2}$ is the flavor singlet field
which has been included as a device to obtain the flavor neutral
propagators in \PQCPT. 
Expanding the Lagrangian to tree-level, one finds that 
mesons composed of a quark $Q_i$ and antiquark $\ol Q_j$ have 
masses given by
\begin{equation}
m_{Q_i Q_j}^2 = \lambda \left[ (m_{Q})_{ii} + (m_Q)_{jj} \right] 
.\end{equation}
The flavor singlet field additionally acquires a mass $m_0^2$. 
In \PQCPT\ (as well as in \CPT), the strong $U(1)_A$ anomaly
generates a mass for the flavor singlet field and hence
$m_0$ can be taken on the order of the chiral symmetry
breaking scale, $m_0 \sim \Lambda_\chi \approx 4 \pi f$. The flavor singlet
field can thus be integrated out. Flavor neutral propagators in \PQCPT, however,
cannot be diagonalized into simple single pole forms~\cite{Sharpe:2001fh}. 
This fact notwithstanding, the results of our calculations 
will not require the explicit form of these flavor neutral propagators.

In writing the above theory of mesons, we have added the effects of 
electromagnetism in the leading-order Lagrangian. The $U(1)$ gauge 
field, $A_\mu$, appears in the action of the covariant derivative, $D_\mu$, 
namely
\begin{equation}
D_\mu \Sigma = \partial_\mu \Sigma + i e A_\mu \left[ \cQ, \Sigma \right]
,\end{equation}
where $\cQ$ is the quark electric charge matrix. To completely specify
how electromagnetism is coupled, we must extend the quark
charges to the partially quenched theory. The choice
\begin{equation}
\cQ = \diag \left( q_u , q_d, q_j, q_l, q_u, q_d \right)
,\end{equation}
with $q_j + q_l \neq 0$, is particularly useful because it retains 
sensitivity to all electromagnetic couplings in the theory
as well as maintains the cancellation of disconnected operator
insertions between the valence and ghost sectors~\cite{Tiburzi:2004mv,Detmold:2005pt}.
Other choices are possible but can be computationally cumbersome in actual lattice simulations.

\subsection{Pion Polarizabilities in Infinite Volume} \label{ppiv}

\begin{figure}[tb]
  \includegraphics[width=0.58\textwidth]{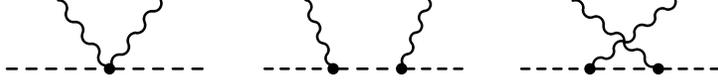}%
  \caption{
Tree-level contributions to the Compton scattering amplitude.
The dashed lines represent mesons, while the wiggly lines represent photons.
Vertices are generated from the leading and next-to-leading order Lagrangian.}
  \label{f:piontree}
\end{figure}

To determine the pion polarizabilities, we calculate the
Compton scattering amplitude for the process $\gamma  \pi  \to \gamma \pi$
using \PQCPT. Contributions to the amplitude are of three types:
tree-level, wavefunction renormalization corrections, and 
one-loop contributions. 
The first contributions arise from tree-level graphs generated 
from local electromagnetic vertices in the effective theory. 
The relevant diagrams have been depicted in Figure~\ref{f:piontree},
and are only non-vanishing for the charged pion.
The first diagram represents the local coupling of two photons 
to the pion. This diagram arises from both the charge-squared operator
contained in the leading-order Lagrangian, as well as from terms
in the next-to-leading order Lagrangian. Specifically in the notation 
of~\cite{Donoghue:1992dd}, the local two-photon, two-pion interactions
are contained in the next-to-leading order terms\footnote{%
Although we use the $SU(3)$ notation for these terms, final 
results depend on the scale-independent combination $\alpha_9 + \alpha_{10}$,
which has the same value in $SU(2)$ as it does in $SU(3)$. 
}
\begin{equation}
\cL 
= 
i \, e \, \alpha_9 \, F_{\mu \nu} \, 
\str \left( 
\cQ D^\mu \Sigma D^\nu \Sigma^\dagger 
+ 
\cQ D^\mu \Sigma^\dagger D^\nu \Sigma
\right)
+
e^2 \alpha_{10}
\, F^2 \str \left( \cQ \Sigma \cQ \Sigma^\dagger \right)
,\end{equation}
where $F_{\mu \nu} = \partial_\mu A_\nu - \partial_\nu A_\mu$ 
is the electromagnetic field-strength tensor. 
In \PQCPT, the low-energy constants $\alpha_9$ and $\alpha_{10}$
have the same numerical values as in \CPT, which can be 
demonstrated by matching. 
The remaining two diagrams in Figure~\ref{f:piontree} 
are Born terms that do not contribute
to the one-pion irreducible Compton amplitude.

\begin{figure}[tb]
  \includegraphics[width=0.4\textwidth]{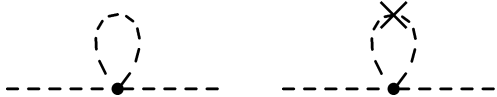}%
  \caption{
  Wavefunction renormalization in \PQCPT. 
Diagram elements are the sames as in Figure~\ref{f:piontree}, 
and the cross denotes the partially quenched hairpin. The vertex 
is generated by the leading-order Lagrangian.}
  \label{f:pionwfn}
\end{figure}

The next contributions are those that arise from the pion 
wavefunction renormalization.  The leading self-energy 
diagrams are depicted in Figure~\ref{f:pionwfn}.
The leading-order diagrams involving the photon coupling to the pion charge 
must be multiplied by the wavefunction renormalization to obtain contributions 
to the Compton amplitude at next-to-leading order. Thus we 
require only the wavefunction renormalization of the 
the charged pion.  Due to fortuitous cancellation in both \PQCPT\ and \QCPT,  
the hairpin diagram, which arises from the double pole structure
of the flavor-neutral propagator, vanishes.

\begin{figure}[tb]
\bigskip
  \includegraphics[width=0.48\textwidth]{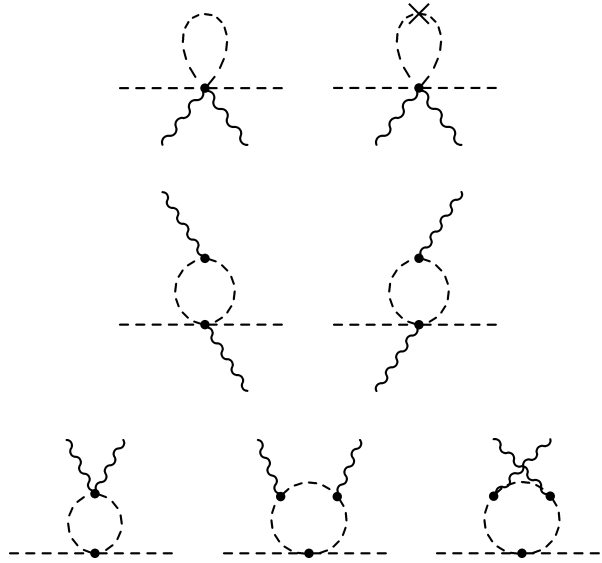}%
  \caption{
One-loop contributions to the Compton scattering amplitude in \PQCPT.
Vertices shown are generated from the leading-order Lagrangian, 
and diagrams depicted are all one-pion irreducible. 
  }
  \label{f:radius}
\end{figure}

\begin{figure}[tb]
\bigskip
  \includegraphics[width=0.58\textwidth]{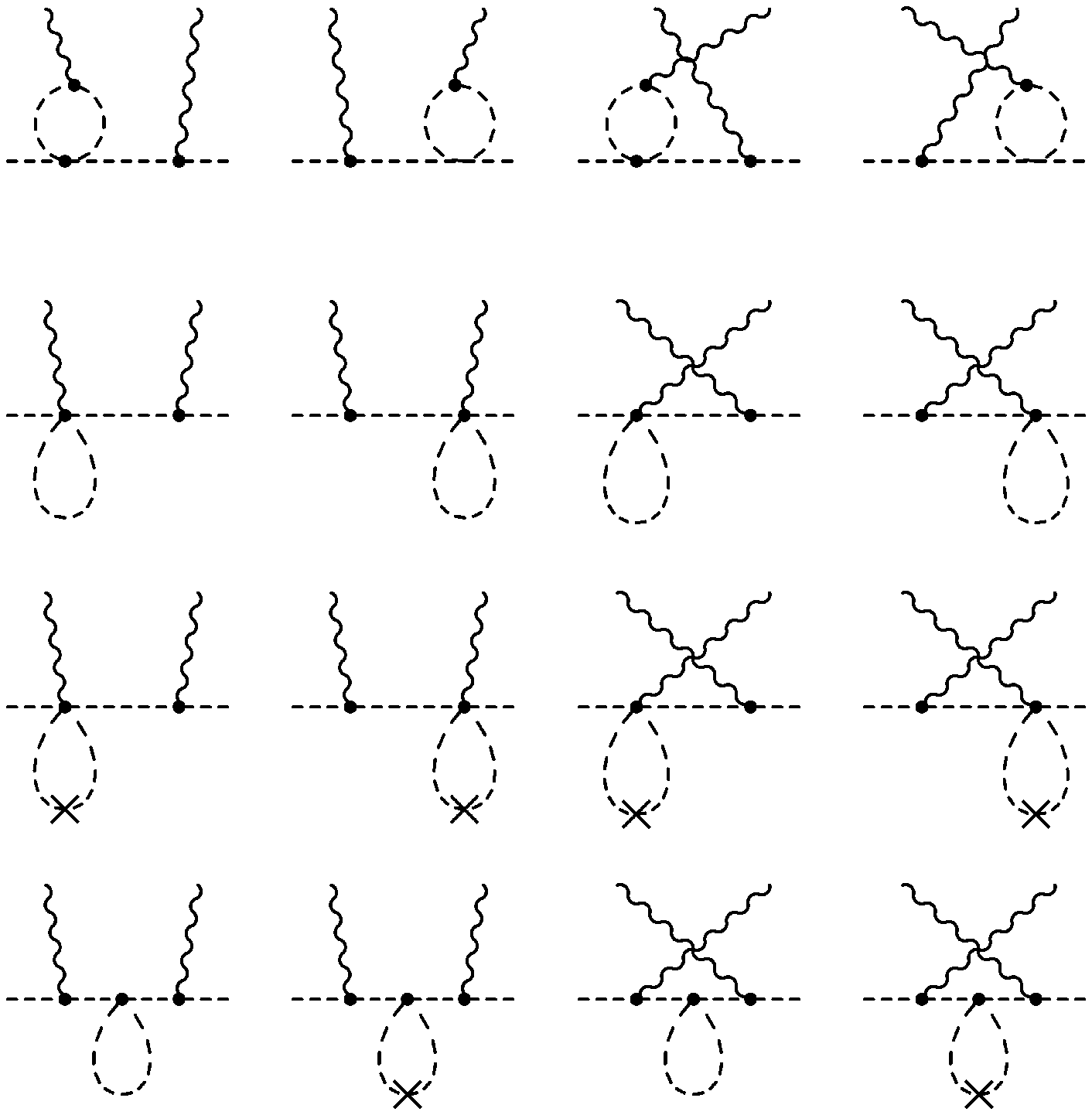}%
  \caption{
One-pion reducible contributions to the Compton scattering amplitude at
one-loop order in \PQCPT.
  }
  \label{f:reducible}
\end{figure}

The remaining contributions to the Compton amplitude arise 
from one-loop diagrams. In Figure~\ref{f:radius}, we display
the diagrams for the one-pion irreducible scattering amplitude.
Contributions from such diagrams lead to chiral corrections to 
the electromagnetic polarizabilities.
For the charged pion, there are additional one-pion reducible pieces in \PQCPT. 
These diagrams are displayed in Figure~\ref{f:reducible}.
The effects of such diagrams in infinite volume, however, are to renormalize
the mass of the intermediate state pion, and to provide the necessary 
cancellations which preserve the charge interaction of the leading Born terms.
The latter cancellations were first worked out explicitly for the 
case of the pion charge radius in \PQCPT\ in~\cite{Arndt:2003ww,Bunton:2006va}.
Assembling the results of Figures~\ref{f:piontree}--\ref{f:reducible},
we can extract the pion polarizabilities using Eq.~\eqref{compamp}
by utilizing Coulomb gauge in the center-of-momentum frame.

At one-loop order, it is well known that $\alpha_E + \beta_M = 0$ for both
charged and neutral pions~\cite{Bijnens:1987dc,Donoghue:1988ee,Holstein:1990qy}. 
We find this remains true to one-loop order in \PQCPT, as well as \QCPT.
This is expected because extending the flavor algebra from 
$SU(2)$ 
to graded Lie algebras cannot 
alter the helicity structure of the Compton amplitude. 
As for the orthogonal combination of polarizabilities, $\alpha_E - \beta_M$, 
we arrive at
\begin{eqnarray} \label{eq:pi0}
\alpha^{\pi^0}_E - \beta^{\pi^0}_M 
&=&
- \frac{2 \alpha_{fs} Q_\pi^2}{3 (4 \pi f)^2 m_\pi}
\\
\alpha^{\pi^\pm}_E - \beta^{\pi^\pm}_M 
&=&
\frac{16 \alpha_{fs} Q_\pi^2 (\alpha_9 + \alpha_{10})}{f^2 m_\pi}
\label{eq:piplus}
,\end{eqnarray}
with 
$Q_\pi = q_u - q_d$. 
These results are the same in \CPT, \PQCPT, and \QCPT, with the 
exception that in \QCPT\ the low-energy constants 
$\a_9 + \a_{10}$, 
and 
$f$ 
have different numerical values. Furthermore 
our \CPT\ result agrees with the literature, see~\cite{Holstein:1990qy}
(being careful to note 
$f = \sqrt{2} f_\pi$). 
In deriving the above result, we remark that the delicate cancellation
between pion loops in the zero frequency limit present in \CPT\ remains in \PQCPT,
and \QCPT. This cancellation is required by the infinite volume gauge invariance 
of the amplitude and reflects that the longest-range coupling to 
the pion is only to the total charge. In this way the Thomson scattering
cross section is produced in these three theories when the zero frequency limit
is taken.

In each theory, there are no local electromagnetic interaction 
terms for the neutral pion in the next-to-leading order Lagrangian. 
Thus there can be no divergences in the polarizabilities of the neutral pion, 
as we found explicitly at one-loop. 
While chiral logarithms are absent for the neutral pion, 
there are finite terms from the loop graphs. 
In fact, the entire pion cloud contribution to the 
neutral pion polarizabilities manages to survive quenching. 
\begin{figure}[tb]
\bigskip
  \includegraphics[width=0.5\textwidth]{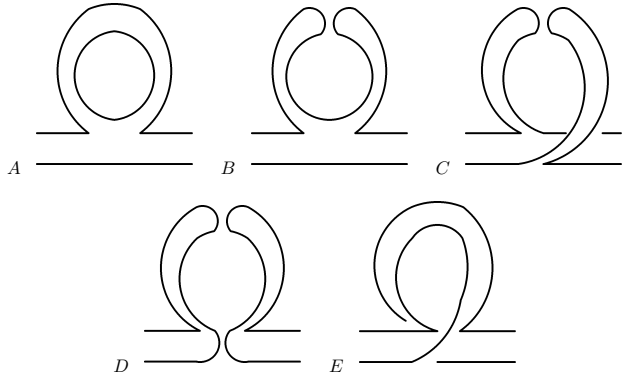}%
  \caption{
Quark-line topologies generated at one-loop order.
Diagrams in the second row contribute only when the
external states are flavor neutral. 
 }
  \label{f:quarkline}
\end{figure}
This is rather surprising, but can be understood by 
considering the quark-line topologies generated at
one-loop order.

The five topologies arising from
the four-pion vertex generated from Eq.~\eqref{eq:L}
are depicted in Figure~\ref{f:quarkline}. 
The topologies in the second row are only possible for
flavor neutral external states, such as the neutral pion.
Let us investigate which topologies can make non-vanishing
contributions to the neutral pion polarizabilities.
Diagram $A$ contains a sea quark loop and thus associated
contributions are proportional to
\begin{equation} \label{eq:PQQ}
\D Q^2 = (q_u - q_j)^2 + (q_u - q_l)^2 + (q_d - q_j)^2 + (q_d - q_l)^2
,\end{equation}
which sums the charge-squared couplings 
from all possible valence-sea loop mesons.
In the isospin limit of 
$SU(4|2)$ 
all such mesons are degenerate with mass 
$m_{ju}$.
The net contribution from topology 
$A$, 
however, vanishes because
contributions from the four-meson vertex with two derivatives
exactly cancel contributions from the four-meson vertex with quark 
mass insertion. 
Terms from all of the meson loop diagrams in Figure~\ref{f:radius}
are required for this delicate cancellation.
As a result, the characteristic factor of 
$\D Q^2$ 
is absent from Eq.~\eqref{eq:pi0}. 
Next, each of the quark line topologies $B$, $C$, and $D$,
require flavor disconnected contributions from 
flavor-neutral meson propagators. 
As flavor neutral mesons are also electrically neutral, 
coupling to the photon eliminates such contributions. 
Indeed looking at Figure~\ref{f:radius}, there is only one possible 
diagram with a hairpin vertex. Direct evaluation shows that this 
contribution vanishes, ruling out the $B$, $C$, and $D$ topologies. 
Therefore the loop contributions to neutral pion polarizabilities
in Eq.~\eqref{eq:pi0} stem entirely from topology $E$. 
As this topology is quark-line connected, the contribution
has the same form regardless of quenching.

Let us examine topology $E$ closer by writing
the pion field in terms of quark basis mesons, $\eta_u$ and $\eta_d$, 
\begin{equation}
\pi^0 
= 
\frac{1}{\sqrt{2}} 
\left( 
\eta_n 
- 
\eta_d 
\right)
.\end{equation}
The diagonal contractions,
$\eta_u$-$\eta_u$
and 
$\eta_d$-$\eta_d$, 
for topology $E$ result 
in an electrically neutral loop meson. 
Only the first diagram of Figure~\ref{f:radius}
could yield the diagonal contractions of topology $E$. 
Close inspection of the Lagrangian shows, however, 
that the four-meson, two-photon vertex with 
four electrically neutral mesons is identically zero. 
Thus neutral pion polarizabilities
stem entirely from topology $E$'s non-diagonal flavor contractions:
$\eta_u$-$\eta_d$, and $\eta_d$-$\eta_u$. 
To one-loop order, the 
neutral pion polarizabilities arise entirely from 
annihilation contractions of lattice QCD correlation functions.

Returning to Eq.~\eqref{eq:piplus}
for the charged pion polarizabilities,
the quark-line picture helps to show why Eq.~\eqref{eq:piplus}
has no loop contributions.
In Figure~\ref{f:quarkline},
the quark-line topologies in the second row are no longer relevant
because the external states are charged.
Furthermore topologies $B$ and $C$ require
hairpins, but the hairpin graphs in Figures~\ref{f:pionwfn} and~\ref{f:radius} vanish.
Loop contributions to charged pion polarizabilities 
can only arise from topology $A$. 
Cancellation of divergent loop contributions from this topology must 
occur in \CPT, \PQCPT, and \QCPT\ because the combination 
$\a_9 + \a_{10}$ 
is renormalization scale independent.
This independence disallows chiral logarithms from loop contributions in \CPT, and \QCPT.
While one can imagine scale invariant combinations of the form 
$\log (m_{ju}^2 / m_{lu}^2)$, say, 
away from the isopsin limit of  $SU(4|2)$, charge-squared couplings
do not allow for loop contributions to alternate in sign, see Eq.~\eqref{eq:PQQ}. 
Thus such logarithms are absent.
While logarithms are not allowed, finite contributions can be 
present. 
As with the neutral pions for topology $A$, however,
the contributions from the four-meson vertex with two derivatives
exactly cancel contributions from the four-meson vertex with quark 
mass insertion. The characteristic factor of $\D Q^2$ 
is consequently absent from from the charged pion polarizabilities, 
Eq.~\eqref{eq:piplus}. 
Thus the accidental cancellation of finite terms in \CPT\
also occurs in \PQCPT, and \QCPT.

As a final comment on the inifinite volume results in  
Eqs.~\eqref{eq:pi0} and \eqref{eq:piplus}, 
the only pion mass dependence in both charged and 
neutral pion polarizabilities arises from the target mass 
$m_\pi$. 
The target mass depends on the valence quark mass.
Both of these statements hold only to one-loop 
order in the chiral expansion.

\section{Compton Tensor in Finite Volume} \label{ppfv}

In finite volume, the pion is already deformed. Thus its ability to polarize in an applied 
electric or magnetic field will differ from that in infinite volume. As finite 
volume modifications to hadron properties are long distance in nature, they can be quite generally 
addressed using \CPT. Lattice simulations are usually carried out in a hypercubic box of
volume 
$L^3 \times \b$, 
where $L$ is the length of the spatial direction, and $\b$ is the length
of the Euclidean time direction. We consider $\b \gg L$ so that there is no effect from the finite 
temporal extent of the lattice. With periodic boundary conditions on the quark fields in each of the spatial directions, 
the momentum modes on the lattice are 
$\bm{p} = 2 \pi \bm{n} / L$, with $\bm{n}$ 
a triplet of integers.
The ordinary power counting for \CPT\
\begin{equation}
|\bm{p}| \lesssim m_\pi  \ll \Lambda_\chi
\end{equation}
can be applied in a box of finite size provided $2 f L \gg 1$ and
$m_\pi L \gtrsim 2 \pi$. These two conditions constitute what is 
called the $p$-regime of chiral perturbation theory. 
The first condition is required in order to use chiral perturbation theory 
at all, while the second condition maintains that pionic zero modes 
remain perturbative~\cite{Gasser:1986vb,Gasser:1987ah}. 
As the power counting in this regime remains the same as in infinite 
volume~\cite{Gasser:1987zq}, 
the same diagrams depicted in Figures~\ref{f:piontree}-\ref{f:reducible}
contribute to the pion Compton tensor. 
It is straightforward to perform the loop calculations in a finite box, 
we merely replace integrals over virtual four-momenta by integrals over 
energy and sums over spatial momentum modes permitted by periodicity.

Consider an observable $X$ calculated in both finite and infinite volume.
Let $X(L)$ denote the value of the observable in finite volume, 
and $X(\infty)$ denote its value in infinite volume. 
The finite and infinite volume theories share exactly the same
ultraviolet divergences, thus the volume effect can be determined
from matching the two theories in the infrared, 
\begin{equation}
X(L) = X(\infty) + \Delta X(L)
.\end{equation} 
The volume effect is given by the matching term $\Delta X(L)$ which
is ultraviolet finite.  A salient feature of such matching is that 
it allows us to retain our infinite volume regularization scheme 
and values of low-energy constants.

Calculating the finite volume matching for the Compton 
scattering amplitude, while straightforward, is quite involved
in practice. 
We first remark that the decomposition of the Compton 
tensor in Eq.~\eqref{compamp} is no longer valid.
That decomposition makes use of the center-of-momentum frame. 
Finite volume results on a torus have only an $S_4$ 
cubic subgroup of the infinite volume $SO(4)$ invariance. 
Results will thus be frame dependent, and hence, to be general, 
we must not make recourse to a particular frame. 
Furthermore, as shown in~\cite{Hu:2007eb}, 
there are more gauge invariant structures
allowed on a torus. Thus more
terms  than shown in Eq.~\eqref{compamp}
are allowed at second order in the field 
strength.

\subsection{Neutral Pion}

Carrying out the finite volume matching
on the neutral pion Compton amplitude
without recourse to a particular frame or gauge, 
we find
\footnotesize
\begin{eqnarray} \label{eq:neutralFV}
\Delta T_{\pi^0}^{\mu \nu} (L) 
&=&
\frac{e^2}{f^2} 
\sum_{\phi} \cC_\phi^{\pi^0}
\Bigg[
- \frac{1}{6} g^{\mu \nu} 
\int_0^1 dx  I_{3/2}(x\bm{r}, m_\phi^2 - x (1-x) r^2)
\notag \\
&&+ 
\d^{\mu 0} \d^{\nu 0}
\Bigg(
\frac{1}{3}
\int_0^1 dx \int_0^{1-x} dy
I_{3/2} (x\bm{k} + y \bm{k'}, m_\phi^2 - xy r^2)
\notag \\
&&
\phantom{indent}
- 
\frac{1}{4} 
\int_0^1 dx \int_0^{1-x} dy 
[(2x-1) \omega + 2 y \omega']
[2 x \omega + (2y-1) \omega']
I_{5/2} (x\bm{k} + y \bm{k'}, m_\phi^2 - xy r^2)
\Bigg)
\notag \\
&&+
\frac{1}{4} 
\d^{\mu 0}\d^{\nu j}
\int_0^1 dx \int_0^{1-x} dy 
[(2 x - 1) \omega + 2 y \omega']
\Big\{
k'^j I_{5/2}(x\bm{k} + y \bm{k'}, m_\phi^2 - xy r^2)
\notag \\
&&\phantom{indent}+ 
2
I_{5/2}^j (x\bm{k} + y \bm{k'}, m_\phi^2 - xy r^2)
\Big\}
\notag \\
&&+
\frac{1}{4} 
\d^{\mu i}\d^{\nu 0}
\int_0^1 dx \int_0^{1-x} dy 
[(2 y - 1) \omega' + 2 x \omega]
\Big\{
k^i I_{5/2}(x\bm{k} + y \bm{k'}, m_\phi^2 - xy r^2)
\notag \\
&&\phantom{indent}+ 
2
I_{5/2}^i (x\bm{k} + y \bm{k'}, m_\phi^2 - xy r^2)
\Big\}
\notag \\
&&-
\frac{1}{4} \d^{\mu i} \d^{\nu j}
\int_0^1 dx \int_0^{1-x} dy 
[
4 I_{5/2}^{ij}(x\bm{k} + y \bm{k'}, m_\phi^2 - xy r^2)
+ 
2 k^i I_{5/2}^j(x\bm{k} + y \bm{k'}, m_\phi^2 - xy r^2)
\notag \\
&&\phantom{indent}+ 
2 k'^j I_{5/2}^i (x\bm{k} + y \bm{k'}, m_\phi^2 - xy r^2)
+ 
k^i k'^j I_{5/2}(x\bm{k} + y \bm{k'}, m_\phi^2 - xy r^2)
]
\Bigg]
\end{eqnarray}
\normalsize
Above we have employed 
$r_\mu = (k-k')_\mu$ 
for the momentum transfer, with
$k_\mu = (\omega, \omega \hat{\bm{k}})$, 
and
$k'_\mu = (\omega', \omega' \hat{\bm{k}}')$
for the photon momenta.
To derive the above result, we have employed the 
reality of initial and final state photons, and
taken their spatial momenta to be quantized.
The finite volume functions 
$I_\b(\bm{\theta},m^2)$, 
$I^i_\b(\bm{\theta},m^2)$, 
and 
$I^{ij}_\b(\bm{\theta},m^2)$ 
are defined in Appendix~\ref{fvf}. 
The coefficient for contributing loop mesons $\cC_\phi^{\pi^0}$
is given by
\begin{equation} \label{eq:neutralpioncoeff}
\cC_\phi^{\pi^0}
=
3 Q_\pi^2 \left(2 m_\pi^2 -   r^2 \right) \delta_{\phi,\pi} 
-
\frac{3}{2} \Delta Q^2  r^2 \delta_{\phi, ju}
.\end{equation}
While we have only given the \PQCPT\ coefficients, 
the \CPT, and \QCPT\ results
can be trivially deduced from Eq.~\eqref{eq:neutralpioncoeff}.
The latter is possible because there are no hairpin contributions. 
At zero frequency, the finite volume Compton amplitude for the neutral 
pion is non-vanishing. This is because the Thomson cross-section is
not protected from renormalization in finite volume~\cite{Hu:2007eb}.

\subsection{Charged Pion}

The charged pion Compton amplitude at finite volume
is even more involved than the neutral result 
as we must determine both reducible
and irreducible contributions. 
To one-loop order, the result is
\footnotesize
\begin{eqnarray}  \label{eq:chargedFV}
\Delta T_{\pi^\pm}^{\mu \nu} (L)
&=&
- \frac{3 e^2 \D Q^2 }{2 f^2} r^2 
\Bigg\{
- \frac{1}{6} g^{\mu \nu} 
\int_0^1  dx I_{3/2}(x\bm{r}, m_{ju}^2 - x (1-x) r^2)
\notag \\
&&+ 
\d^{\mu 0} \d^{\nu 0}
\int_0^1 dx \int_0^{1-x} dy
\Bigg[
\frac{1}{3}
I_{3/2} (x\bm{k} + y \bm{k'}, m_{ju}^2 - xy r^2)
\notag \\
&&
\phantom{indent}
- 
\frac{1}{4}  
[(2x-1) \omega + 2 y \omega']
[2 x \omega + (2y-1) \omega']
I_{5/2} (x\bm{k} + y \bm{k'}, m_{ju}^2 - xy r^2)
\Bigg]
\notag \\
&&+
\frac{1}{4} 
\d^{\mu 0}\d^{\nu j}
\int_0^1 dx \int_0^{1-x} dy
[(2 x - 1) \omega + 2 y \omega']
\notag \\
&&\phantom{indent} \times 
\Big[
k'^j I_{5/2}(x\bm{k} + y \bm{k'}, m_{ju}^2 - xy r^2)
+ 
2
I_{5/2}^j (x\bm{k} + y \bm{k'}, m_{ju}^2 - xy r^2)
\Big]
\notag \\
&&+
\frac{1}{4} 
\d^{\mu i}\d^{\nu 0}
\int_0^1 dx \int_0^{1-x} dy
[(2 y - 1) \omega' + 2 x \omega]
\notag \\
&&\phantom{indent} \times 
\Big[
k^i I_{5/2}(x\bm{k} + y \bm{k'}, m_{ju}^2 - xy r^2)
+ 
2
I_{5/2}^i (x\bm{k} + y \bm{k'}, m_{ju}^2 - xy r^2)
\Big]
\notag \\
&&-
\frac{1}{4} \d^{\mu i} \d^{\nu j}
\int_0^1 dx \int_0^{1-x} dy
\Big[
4 I_{5/2}^{ij}(x\bm{k} + y \bm{k'}, m_{ju}^2 - xy r^2)
+ 
2 k^i I_{5/2}^j(x\bm{k} + y \bm{k'}, m_{ju}^2 - xy r^2)
\notag \\
&& \phantom{indent}
+ 
2 k'^j I_{5/2}^i (x\bm{k} + y \bm{k'}, m_{ju}^2 - xy r^2)
+ 
k^i k'^j I_{5/2}(x\bm{k} + y \bm{k'}, m_{ju}^2 - xy r^2)
\Big]
\Bigg\}
\notag \\
&&
- \frac{2 e^2 Q_\pi^2}{f^2}
\Bigg[
2  g^{\mu \nu}
- 
\frac{(2 P+k)^\mu (P+P'+k)^\nu}{(P+k)^2 - m_\pi^2}
- 
\frac{(2 P - k')^\nu (P + P' - k')^\mu}{(P - k')^2 - m_\pi^2}
\Bigg]
I_{1/2}(m_{ju}^2)
\notag \\
&&
- e^2 Q_{\pi}^2
\Bigg[
\frac{I^\mu(P, P+k) (P+P'+k)^\nu}{(P+k)^2 - m_\pi^2} 
+ 
\frac{(2 P + k)^\mu I^\nu(P+k, P')}{(P+k)^2 - m_\pi^2}
\notag 
\\
&&
\phantom{spa}+ 
\frac{I^\mu(P-k', P') (2 P - k')^\nu}{(P - k')^2 - m_\pi^2}
+
\frac{(P+P' - k')^\mu I^\nu(P, P-k')}{(P - k')^2 - m_\pi^2}
\Bigg]
\notag \\
&&+
\frac{e^2 Q_\pi^2}{f^2}
\Bigg[
\d^{\mu 0} \d^{\nu 0}
\int_0^1 dx
\Big[ 
2 I_{1/2}(x \bm{k}, m_{ju}^2) 
+ 
2 I_{1/2}(x \bm{k'}, m_{ju}^2)
\notag \\
&&\phantom{indentation}
-
x (2 x - 1)
\left(
\o^2 
I_{3/2}(x \bm{k}, m_{ju}^2) 
+ 
\o'^2
I_{3/2}(x \bm{k'}, m_{ju}^2) 
\right) \Big]
\notag \\
&&+
\d^{\mu 0} \d^{\nu j}
\int_0^1 d x \,
\left[ 
x \o' k'^j I_{3/2}(x \bm{k'}, m_{ju}^2)
+
2 
x \o'
I^j_{3/2}(x \bm{k'}, m_{ju}^2)
+ 
(2 x -1) \o I^j_{3/2}(x \bm{k}, m_{ju}^2)
\right]
\notag \\
&&+
\d^{\mu i} \d^{\nu 0}
\int_0^1 dx  \,
\left[ 
x \o k^i I_{3/2}(x \bm{k}, m_{ju}^2)
+
2 x \o I^j_{3/2}(x \bm{k}, m_{ju}^2)
+ 
(2 x -1) \o' I^i_{3/2}(x \bm{k'}, m_{ju}^2)
\right]
\notag\\
&&-
\d^{\mu i} \d^{\nu j}
\int_0^1 dx
\Big[
2  I^{ij}_{3/2}(x \bm{k}, m_{ju}^2)
+
2  I^{ij}_{3/2}(x \bm{k'}, m_{ju}^2)
\notag\\
&&\phantom{indent}+
k^i I^{j}_{3/2}(x \bm{k}, m_{ju}^2)
+
k'^j I^{i}_{3/2}(x \bm{k'}, m_{ju}^2)
\Big]
\Bigg\}
.\end{eqnarray}
\normalsize
In the above result, 
the initial (final) pion momentum has been denoted by 
$P$ ($P'$). 
We have employed an abbreviation for the finite volume
pion-photon vertex function, 
$I^\mu(P_2,P_1)$, 
which arises from the one-pion reducible diagrams and is given by
\footnotesize
\begin{eqnarray}
I^\mu(P_2,P_1)
&=&
\frac{1}{f^2}
\d^{\mu 0}
\Bigg\{
(P_2 + P_1)^0 
\Big[
\int_0^1 dx  
I_{1/2} (x \bm{\D}, m^2_{ju} - x (1-x) \D^2)
\notag \\
&&\phantom{indentindent}
-\frac{1}{2} (\D^0)^2 
\int_0^1 dx \, x (2 x - 1) I_{3/2} (x \bm{\D}, m^2_{ju} - x (1-x) \D^2)
\Big]
\notag \\
&&\phantom{indent}
+ 
\frac{1}{2} \Delta^0
\int_0^1  dx (1-2x) 
(\bm{P}_2 + \bm{P}_1)
\cdot
\bm{I}_{3/2} (x \bm{\D}, m^2_{ju} - x (1-x) \D^2)
\Bigg\}
\notag \\
&&+
\frac{1}{f^2}
\d^{\mu j}
\Bigg\{
(P_2 + P_1)^i 
\Big[
\int_0^1 dx
I^{ij}_{3/2} (x \bm{\D}, m^2_{ju} - x (1-x) \D^2)
\notag \\
&&\phantom{indentindent}
+
\frac{1}{2} \D^j 
\int_0^1 dx I^{i}_{3/2}(x \bm{\D}, m^2_{ju} - x (1-x) \D^2)
\Big]
\notag\\
&&\phantom{indent}
+
\D^0 (P_2 + P_1)^0
\Big[
\int_0^1  dx \, x 
\Big[
I^j_{3/2}(x \bm{\D}, m^2_{ju} - x (1-x) \D^2) 
\notag \\
&&\phantom{indentindent}
+ 
\frac{1}{2} \D^j 
\int_0^1 dx I_{3/2}(x \bm{\D}, m^2_{ju} - x (1-x) \D^2)
\Big]
\Bigg\}
,\end{eqnarray}
\normalsize
with 
$\Delta^\mu = (P_2- P_1)^\mu$. 
At zero frequency, we recover the results of~\cite{Hu:2007eb}.
Specifically from the one-pion reducible terms, we see
that the current is renormalized. This is possible at finite 
volume because of gauge invariant zero-mode interactions.

\subsection{Discussion of Finite Volume Results}

With Eqs.~\eqref{eq:neutralFV} and \eqref{eq:chargedFV}, 
we have deduced the finite volume modification
to the pion Compton scattering tensor. 
These results show explicitly broken $SO(4)$
invariance as well as additional structures
not anticipated by infinite volume gauge invariance.
The finite volume modifications  can be directly utilized if 
two-current, two-pion correlation functions are calculated on
the lattice. One merely removes the finite volume effects determined 
above to isolate the infinite volume physics. 
Such lattice calculations of the Compton tensor are, 
however, prohibitively expensive time wise, and will not 
be performed in the foreseeable future. A practical alternative 
to these calculations is provided by the background field method. 
In this approach, a classical electromagnetic field
is gauged into the QCD action.\footnote{%
Implementing this method currently suffers
the need to quench effects of the background field. 
In principle, there is no impediment to coupling a suitably 
weak background field to sea quarks other than time cost.   
}
One then studies the external field dependence of
correlation functions to deduce electromagnetic 
observables.  For example, at infinite volume
the energy of a neutral pion in a weak external 
electric field is 
\begin{equation}
E_\pi(\bm{p}=\bm{0}) 
= 
m_\pi 
-
 \frac{1}{2} \alpha^\pi_E \bm{E}^2
 + 
 \cO(\bm{E}^4)
.\end{equation}
Thus by measuring the quadratic energy shift in the
external field strength $|\bm{E}|$, one can deduce
the electric polarizability.  
A practical question is then how 
to deduce volume corrections to polarizabilities
determined from background field methods.
Given the relation of the infinite volume Compton 
tensor to the polarizabilities, one might suspect
that the finite volume Compton tensor in
Eqs.~\eqref{eq:neutralFV} and \eqref{eq:chargedFV} 
contains the finite volume corrections to the polarizabilities.
We argue that the finite volume Compton tensor
has no relevance to volume effects in background field 
methods. At finite volume, there is no longer a 
discernible relation between
polarizabilities and the Compton tensor.

An analysis of  finite volume effects for nucleon
polarizabilities for background field methods
derived from the Compton tensor, however, 
was presented in \cite{Detmold:2006vu}.
That analysis employed the Breit frame
decomposition of the nucleon Compton tensor in Coulomb 
gauge.  Such results surely cannot be utilized for background
field calculations because such calculations are typically done
in the rest frame. The finite volume modifications
derived, moreover, are polluted by subtle effects from the gauge field 
due to the nature of gauge invariance on a torus. 
These effects have nothing to do with polarizabilities. 
For example, in the center-of-momentum frame, where 
$k^0 = k'^0 = \o$, 
we may encounter a term in the amplitude of the form
\begin{equation} \label{eq:amp}
\cM  
= 
\ldots 
+
\frac{1}{2}
\o^2 \alpha(L)  \, 
\bm{\varepsilon}'^* 
\cdot 
\bm{\varepsilon} 
+ 
\ldots 
,\end{equation}
and be tempted to conclude that 
$\alpha(L)$ 
is a finite volume correction
to the electric polarizability.
In a general frame, however, this term could stem from any 
combination of 
$\o^2$, 
$\o'^2$,
and 
$\o \o'$ structures. 
In infinite volume only the last term is allowed by gauge invariance, specifically 
by an operator $\propto \bm{E}^2$ with
a coefficient proportional to the electric polarizability. 
In finite volume, however,  the additional structures $\o^2$
and $\o'^2$ are allowed. They stem from single-particle 
effective theory operators of the form
\begin{equation} \label{eq:zeroEop}
\cL = \frac{i}{2} \ol \alpha(L)  \, \bm{W}^{(-)} \cdot \frac{\partial \bm{E} }{\partial t}  \,  \,   \tr ( Q^2 \Phi^2 )
,\end{equation}
for example, where 
$W_i^{(-)}$ 
is the negative parity part of the zero-mode Wilson line 
$W_i$, 
given by
\begin{equation}
W_i = \cP_{\bm{0}} \cW_i \cP^\dagger_{\bm{0}}
,\end{equation}
with the Wilson line $\cW_i$ as
\begin{equation}
\cW_i
=
\exp \left( \frac{i e}{3} \oint dx_i A_i \right) 
,\end{equation}
and $\cP_{\bm{0}}$ as the zero-mode projection operator. 
The operator in Eq.~\eqref{eq:zeroEop}
respects $C$, $P$, and $T$, as well as the 
cubic symmetry of the torus.
Furthermore it is gauge invariant
because the zero mode has a periodicity constraint
under gauge transformations, see~\cite{Hu:2007eb}.
From Eq.~\eqref{eq:amp}, we cannot deduce
that $\alpha(L)$ is a finite volume correction to the electric
polarizability.  
In general, one must work in an arbitrary frame 
to disentangle the zero-mode electric coupling
in Eq.~\eqref{eq:zeroEop} from the electric
polarizability. An analogous situation exists
for magnetic interactions, 
because the operator, 
\begin{equation}
\bm{\nabla} \cdot \left( \bm{W}^{(-)} \times \bm{B} \right)  \tr ( Q^2 \Phi^2 )
,\end{equation}
for example, is allowed by symmetries.

The frame and gauge dependence notwithstanding, 
finite volume modifications to polarizabilities were determined in~\cite{Detmold:2006vu} 
from Taylor series expanding the Compton amplitude in 
photon frequency. That procedure is also invalid as we 
now demonstrate. 
For simplicity, consider the following finite volume difference
function, $I_{1/2}(\bm{k}, m^2)$, where $\bm{k}$ is an external
photon momentum. 
To determine finite volume corrections to polarizabilities 
stemming from this term, we perform a Taylor series 
expansion in the external momentum and arrive at
\begin{equation} \label{eq:expand}
I_{1/2}(\bm{k}, m^2) 
= 
I_{1/2} (\bm{0}, m^2) 
- 
\frac{1}{2} \bm{k}^2
m^2 
I_{5/2} (\bm{0}, m^2)
+ 
\cO(\bm{k}^4)
.\end{equation}
If we were interested in determining a hypothetical 
polarizability $X$
entering the amplitude in the form 
\begin{equation}
\cM  = \ldots + \frac{1}{2} \bm{k}^2 X + \ldots
,\end{equation}
then we would be tempted to conclude that
the finite volume effect $\D X(L)$ is given by
\begin{equation}
\D X(L) = - m^2 I_{5/2}(\bm{0}, m^2)
.\end{equation}
Because the external momentum is itself
quantized, instead of Eq.~\eqref{eq:expand} we actually have
the exact relation
\begin{equation} \label{eq:shift}
I_{1/2}(\bm{k}, m^2) 
= 
I_{1/2} (\bm{0}, m^2) 
.\end{equation}
This follows trivially from re-indexing the summation over loop momentum 
modes, or from the periodicity of the elliptic-theta function, see Appendix~\ref{fvf}.
Hence the volume effect for our example is actually $\D X(L) = 0$.
The reason for this discrepancy is a poorly convergent series expansion.\footnote{% 
Another difference between Eqs.~\eqref{eq:expand} and \eqref{eq:shift} is that
the order of summation and differentiation has been interchanged. 
One can easily show, however,
that the summation over modes is uniformly convergent by using the  Weierstrass 
$M$-test. 
}
Naively the expansion is in $\bm{k}^2 = 4  \pi^2  \bm{n}^2 / L^2$, and thus for
large enough box size the finite volume effect should be well approximated by the first few terms
in the Taylor series.  This is not the case. Because higher-order terms have more derivatives, 
these contributions effectively have more propagators and hence more sensitivity to the infrared.
While we would expect the second term in Eq.~\eqref{eq:expand} to be $1/L^2$ suppressed relative to the first term, the asymptotics show that the volume effect is 
$L^2$ enhanced
\begin{equation}
\lim_{L \to \infty}
m^2 I_{5/2} (\bm{0}, m^2) / I_{1/2}(\bm{0}, m^2)
=
\frac{1}{3} L^2
.\end{equation}
The series expansion continues in this fashion: all terms are order one. 
We can see the same effect more directly by expressing the finite volume difference
in terms of the elliptic-theta function, namely
\begin{equation}
I_{1/2}(\bm{k}, m^2) 
= 
\frac{1}{\pi^2 L^2}
\int_0^\infty d \lambda
e^{- m^2 L^2 / 4 \lambda}
\Big[
\vartheta_3 ( \pi n, e^{-\lambda})
\vartheta_3( 0, e^{-\lambda})^2
-
1
\Big]
,\end{equation}
for the choice $\bm{k} = ( 2 \pi n / L , 0 , 0 )$. 
A series expansion in $\bm{k}$ is thus effectively the same as 
expanding in $\pi n$.

Returning to Eqs.~\eqref{eq:neutralFV} and \eqref{eq:chargedFV}, 
we must ascertain whether we can make sense of a series expansion 
in frequency for the Compton tensor. 
Terms of the form 
\begin{equation}
r^2  \int_0^1 dx I_{3/2} (x \bm{r}, m^2 - x(1-x) r^2) 
,\end{equation}
for example, can be plausibly expanded to second order 
because this requires only evaluation of the finite volume 
function at  $\bm{r} = 0$. This was the logic employed
in~\cite{Beane:2004tw} to deduce finite volume corrections 
to the nucleon magnetic moment.  
As $\bm{r}$ is not continuous, however, one cannot 
deduce the small momentum behavior of this term from 
evaluation at $\bm{r} = 0$, nor can one deduce the 
small momentum behavior from Taylor series expanding
other terms like
\begin{equation}
 \int_0^1 dx I_{3/2} (x \bm{r}, m^2 - x(1-x) r^2) 
,\end{equation}
in the Compton tensor, for example.  
Series expanding in $\bm{r} L = 2 \pi \bm{n}$ is nonsense
no matter the size of the box length $L$.\footnote{%
There is a putative improvement in the convergence of the last 
term due to the integral over the Feynman parameter. 
The $L$ scaling of terms in the expansion, however, is unchanged.
}

At finite volume, one must treat the terms in the amplitude as form factors 
in $\omega L$.  Thus for electromagnetic form factors at finite volume, for example, 
volume corrections to electromagnetic moments cannot be deduced. 
Similarly we are unable to use our results for the finite volume modification
of the Compton amplitude to deduce corrections to the pion polarizabilities. 
At second order in the field strength there are a myriad of new terms 
allowed by the less restrictive symmetries on a torus: cubic invariance 
and periodic zero-mode gauge invariance.
Furthermore a small frequency expansion at finite volume does not make sense 
for quantized momenta. Said another way, periodic gauge potentials on a torus
do not lead to electromagnetic multipole expansions.

\section{Summary}  \label{summy}

Above we have investigated chiral and volume corrections 
to pion Compton scattering using \CPT, \PQCPT, and \QCPT. 
In infinte volume, straightforward calculation of the Compton
amplitude allows us to determine charged and neutral
pion polarizabilities in these theories. Due to fortuitous cancellation
there is no dependence on the sea quark masses, or sea quark 
charges at one-loop order in the chiral expansion. 
The Compton tensor itself does not have any quark mass dependence 
at this order. Consequently the quark mass dependence 
of the derived polarizabilities stems from a kinematical prefactor of the inverse target mass. 
As this valence pion mass is relatively inexpensive
to dial, the chiral singularity should be discernible from lattice data
at light quark masses.
Thus as the chiral regime is approached, 
one can use the lattice as a diagnostic
tool to study the chiral behavior of pion polarizabilities.
This can be done most easily for the charged pion.
Whereas for the neutral pion, we demonstrated that
the polarizabilities at one-loop order stem entirely from 
annihilation contractions which are notoriously difficult to
calculate on the lattice.

When accounting for finite volume effects, however, the situation 
becomes more complicated. 
Breaking of $SO(4)$ invariance and the nature of gauge 
invariance on a torus lead to considerably complicated structure 
for the Compton tensor. Sea quark charge and mass dependence 
enter in the one-loop finite volume effects. One cannot unambiguously
determine the volume effects for the polarizabilities from the Compton
tensor because the Taylor series expansion in quantized momentum
is poorly convergent.  What was in infinte volume a series expansion 
in $\omega / m_\pi \ll 1$ that lead to the polarizabilities, now is accompanied by an ill-defined 
expansion in $\omega L \sim 1$ at finite volume. 
This means that even at low energies, the finite volume Compton
amplitude is a form factor in $\o L$. 
Consequently connection of our finite volume 
results to background field lattice calculations is not possible. 
Similarly finite volume corrections to electromagnetic moments 
cannot be deduced from momentum expanding finite volume form factors.
Further investigation is required to determine volume corrections relevant
for observables determined with background field methods.

\begin{acknowledgments}
We thank W.~Detmold, T. Mehen, and A. Walker-Loud for various discussions.
This work is supported in part by the U.S.~Dept.~of Energy,
Grant Nos.~DE-FG02-05ER-41368-0 (J.H.~and B.C.T.), 
DE-FG02-93ER-40762 (B.C.T), 
and by the Schweizerischer Nationalfonds (F.-J.J.). 
\end{acknowledgments}

\appendix

\section{Quenched \CPT} \label{quenched}

Here we give the relevant details needed in the calculation of quenched pion polarizabilities.
In quenched QCD, contributions from sea quarks are completely neglected.
In a quenched theory of two flavors $u$ and $d$, we additionally have 
two ghost quarks $\tilde{u}$ and $\tilde{d}$. The mass matrix is now
\begin{equation}
m_Q = \diag \left( m_u, m_d, m_u, m_d \right)
,\end{equation}
where the final two entries are the masses of the ghost quarks.
These equal mass ghost quarks are necessitated
so that path integral determinants for the valence quarks are 
exactly canceled by those from the ghosts.
The symmetry breaking pattern in \QCPT\ schematically takes the form
$U(2|2)_L \otimes U(2|2)_R \to U(2|2)_V$ because there
is no axial anomaly in quenched QCD. The coset field $\Sigma$
is hence a $U(2|2)$ matrix and the singlet component cannot 
be integrated out.
The dynamics of the pseudo-Goldstone modes is described at leading-order by 
the \QCPT\ Lagrangian
\begin{equation}
\cL 
= 
\frac{f^2}{8} 
\str \left( D_\mu \Sigma^\dagger D^\mu \Sigma \right)
+
\lambda \frac{f^2}{4} 
\str \left( m_Q^\dagger \Sigma + \Sigma^\dagger m_Q \right)
+
\alpha_\Phi D_\mu \Phi_0 D^\mu \Phi_0 - m_0^2 \Phi_0^2
.\end{equation}
While propagators for flavor-neutral mesons have double poles,
these are not encountered explicitly in expressions for
the pion polarizabilities at next-to-leading order. 
Quenched observables are in general unrelated to their
unquenched counterparts, for example, the constants $\alpha_\Phi$ and
$m_0$ have no analogs in \CPT, moreover, arbitrary polynomial functions
of $\Phi_0^2$ can multiply any term in the Lagrangian and 
the low-energy constants in the quenched chiral Lagrangian above result
from treating these polynomial terms in mean-field approximation.

For the quenched electric charge matrix of the quarks, we must have
\begin{equation}
\cQ = \diag \left( q_u , q_d, q_u, q_d \right)
,\end{equation}
for which the condition $\str \cQ = 0$ is unavoidable. In general there are fewer
local electromagnetic terms in \QCPT\ as compared to \CPT. At next-to-leading
order, however, both the $\alpha_9$ and $\alpha_{10}$ terms remain. 
We must keep in mind that the numerical values of these
coefficients are unrelated to their values in \CPT.
Calculation of the pion polarizabilities then proceeds analogously
to the partially quenched case. Results for the quenched polarizabilities
have been given for infinite and finite volume in the main text.

\section{Finite Volume Functions} \label{fvf}

Above we have determined the finite volume modification
to the Compton scattering tensor. In this Appendix, 
we give explicit formulae for the finite volume functions
used to express finite volume differences. 
We use similar notation for these
functions as~\cite{Sachrajda:2004mi,Tiburzi:2006px}, 
where further discussion can be found.

In evaluating a Feynman diagram in finite volume, 
the loop integral is converted into a sum over the allowed
Fourier modes in a periodic box. The difference of this 
sum and the infinite volume result is the finite volume effect.
As is customary,
we treat the length of the time direction as infinite.
All finite volume differences with momentum insertion 
can be cast in terms of the function 
$I_\beta^{i_1 \cdots i_j} (\bm{\theta},m^2)$, 
defined by
\begin{equation}
I^{i_1 \cdots i_j}_\b (\bm{\theta}, m^2)
=
\frac{1}{L^3} 
\sum_{\bm{n}} 
\frac{q^{i_1} \cdots q^{i_j}}{[(\bm{q} + \bm{\theta})^2 + m^2 ]^\b}
- 
\int \frac{d \bm{q}}{(2 \pi)^3}
\frac{q^{i_1} \cdots q^{i_j}}{[(\bm{q} + \bm{\theta})^2 + m^2 ]^\b}
,\end{equation}
where the sum on $\bm{n}$ is over triplets of integers,
and the loop momentum modes are quantized as $\bm{q} = 2 \pi \bm{n} / L$
in a periodic box.
While a general expression for the exponentially convergent form of 
$I_\beta^{i_1 \cdots i_j} (\bm{\theta},m^2)$
exists, it is easiest merely to cite the required cases for our work. These are
\begin{eqnarray}
I_\b (\bm{\theta}, m^2)
&=&
\frac{(L^2 / 4)^{\b - 3/2}}{ (4 \pi)^{3/2} \Gamma(\beta)}
\int_0^\infty d\lambda \,
\lambda^{1/2 - \b} e^{- m^2 L^2 / 4 \lambda} 
\left[
\prod_{j=1}^3
\vartheta_3( \theta_j L / 2 , e^{-\lambda}) - 1
\right]
\\
I_\b^{i_1} (\bm{\theta}, m^2)
&=&
-\frac{1}{2 (\b - 1)}
\frac{d}{d\theta^{i_1}}
I_\b(\bm{\theta},m^2)
- 
\theta^{i_1} I_\b
(\bm{\theta}, m^2)
\\
I_\b^{i_1 i_2} (\bm{\theta}, m^2)
&=&
\frac{1}{4 (\b - 2)(\b - 1)}
\frac{d^2}{d\theta^{i_1} d\theta^{i_2}}
I_{\b-2}(\bm{\theta},m^2)
+
\frac{1}{2 (\b - 1)}
\d^{i_1 i_2} I_{\b - 1} (\bm{\theta},m^2)
\notag \\
&&\phantom{indent}
-
\theta^{i_1} I^{i_2}_\b(\bm{\theta}, m^2)
- 
\theta^{i_2} I^{i_1}_\b(\bm{\theta}, m^2)
- 
\theta^{i_1} \theta^{i_2} I_\b(\bm{\theta}, m^2)
\end{eqnarray}
where $\vartheta_3 (z, q)$ is a Jacobi elliptic-theta
function of the third kind, see, e.g.~\cite{Zinn-Justin:1996cy}.

\bibliography{hb}

\end{document}